\documentstyle[sprocl,epsfig,amssymb,graphicx,multicol]{article}

\begin{document}
\def\integ{\int_{-1/2}^{1/2}}

\title{A NEW BASIS FUNCTION APPROACH TO 'T~HOOFT EQUATION}
\author{O. ABE}
\address{Laboratory of Physics\\ 
Asahikawa Campus, Hokkaido University of Education\\
9 Hokumoncho, Asahikawa 070-8621, Japan\\
E-mail: osamu@asa.hokkyodai.ac.jp}

\maketitle
\abstracts{We present the new basis functions to investigate the 
't~Hooft equation, the lowest order mesonic Light-Front Tamm-Dancoff 
equation for $\rm SU(N_C)$ gauge theories. We find the 
wave function can be well approximated by new basis functions and 
obtain an analytic formula for the mass of the 
lightest bound state. Its value is consistent with the precedent
results.
}

\section{Introduction}
It is expected that light-front (LF) quantization provides an 
powerful tool for studying many-body relativistic field 
theories~\cite{brodsky98,burkardt}. The bare vacuum is equal to the 
physical vacuum in the LF coordinate, since all constituents must 
have non-negative longitudinal momentum. 
This simple structure of the true vacuum 
enables us to avoid the insuperable problems which appeared in the 
Tamm-Dancoff (TD) approximation~\cite{tammdancoff} in the equal time 
frame. Therefore, the TD approximation is 
commonly used in the context of the LF quantization~\cite{lftd}.

Many authors~\cite{mo}${}^-$\cite{sugihara} have developed the 
effectual techniques  for solving LFTD equations in several models.
Bergknoff~\cite{bergknoff} first applied LFTD 
approximation to the massive Schwinger model model~\cite{coleman}, 
which is the extension of the simplest 
(1+1)-dimensional QED${}_2$~\cite{schwinger}. 
He obtained the so-called Bergknoff equation, which is the light 
front Einstein-Schr\"odinger equation truncated to one 
fermion-antifermion pair. Mo and Perry~\cite{mo} presented a 
prevailing method to handle the ground state and the excited state 
in the massive Schwinger model. 
They concluded that the Jacobi polynomials are 
appropriate basis functions to study the massive Schwinger model. 
Harada and his coworker~\cite{harada94} 
investigated the massive Schwinger model with $\rm SU(2)$ flavor 
symmetry, including up to four fermion sectors. They applied of 
simpler basis functions, which are essentially equivalent to the 
Jacobi polynomials, to the model.
Sugihara and collaborators~\cite{sugihara} studied  
2-dimensional $\rm SU(N_C)$ Quantum 
ChromoDynamics(QCD)~\cite{thooft}, 
including four fermion sectors, by means of the basis functions of 
Harada {\it et al}.

Although excellent papers exist concerning massless and massive 
Schwinger models and 2-dimensional 
QCD~\cite{huang}${}^-$\cite{abe99}, 
including excited states, 
it is worth analyzing the ``'t~Hooft'' equation, the lowest order
LFTD equation in $\rm SU(N_C)$ gauge theory. 
This is because there is 
a mathematical interest in the basis function method. There is no 
mathematical evidence that the wave function can be expanded in 
terms of the conventional basis function. Contrarily, 
there is an evidence that the conventional method 
breaks down if we try to improve the approximation. We would 
therefore like to improve the basis functions so as to avoid such 
difficulties.
\section{Conventional Basis function Method}
The 't~Hooft equation for two dimensional $\rm SU(N_C)$ gauge theory 
is given in the form 
\begin{eqnarray}
M^2\Phi (x) &=& \int_{-1/2}^{1/2}dy H(x,y)\Phi(y)\nonumber\\
&\equiv&{4(m^2-1)\over 1-4x^2}\Phi (x)-\wp\int_{-1/2}^{1/2}dy 
   {\Phi (y)\over (y-x)^2},
\quad
-1/2\le x\le 1/2,
\label{thooft}
\end{eqnarray}
where $\Phi$ is a wave function of a bound state, $M$ denotes the
normalized meson mass, $m$ stands for the normalized (anti-)quark 
mass and $\wp$ denotes the finite part integral.
In Eq. (\ref{thooft}), we shifted the variable $x$ total amount of 
$-1/2$ compared with the variable in  Refs.~\cite{bergknoff,thooft}, 
in order to show the symmetry 
of the wave function transparently.

According to Harada and collaborators~\cite{harada94},
one can expect that the wave function could be expanded as follows: 
\begin{eqnarray}
\Phi (x)=\lim_{N\to\infty}
             \sum_{j=0}^{N} a_{j}  (1-4x^2)^{\beta +j}.
\label{expansion}
\end{eqnarray}
The exponent $\beta$ and the normalized quark mass $m$ are related 
to each other 
by the  equation~\cite{bergknoff,thooft}
\begin{eqnarray}
(m^2-1) + \beta\pi\cot\beta\pi =0.
\label{mass}
\end{eqnarray}
The authors of references~\cite{harada94,sugihara,bergknoff} 
adopted the positive smallest solution $\beta_0(m)$ of 
Eq. (\ref{mass}) as $\beta$ in Eq. (\ref{expansion}). 

Mo and Perry, and Harada and his collaborators presented effective 
way to determine the coefficients $a_j$'s. We will briefly reproduce 
their procedures. By the use of the expansion in 
Eq. (\ref{expansion}) truncated to given finite number $N$ for the 
wave function $\Phi $, we multiply both sides of Eq. (\ref{thooft}) 
by $(1-4x^2)^{\beta +i}$ and integrate them over x, then we obtain 
\begin{eqnarray}
M^2 {\widehat N}\vec{a}={\widehat H}\vec{a},\quad 
\vec{a}={}^t[a_0,a_1,\cdots ,a_{n-1}].
\label{eigen1}
\end{eqnarray}

In order to obtain eigenvalues of the generalized eigenvalue 
equation, we have to solve the eigenvalue problem for norm matrix 
$\widehat N$, first, {\it i.e.},

\noindent
\begin{minipage}{0.5\textwidth}
\begin{eqnarray} 
{\widehat N} \vec{v}_i =\lambda_i \vec{v}_i.
\end{eqnarray}
Next, we introduce a transformation matrix $\widehat W$ by
\begin{eqnarray}
{\widehat W}=\left[{\vec{v}_1\over ||\vec{v}_1||\sqrt{\lambda_1}},
    \cdots ,{\vec{v}_n\over ||\vec{v}_n||\sqrt{\lambda_n}}\right].
\end{eqnarray}
Then, we can transform Eq. (\ref{eigen1}) into a usual eigenvalue 
problem of the form
\begin{eqnarray}
M^2 \vec{b}={}^t{\widehat W}{\widehat H}{\widehat W}\vec{b},
\quad \vec{a}={\widehat W}\vec{b}.
\label{eigen2}
\end{eqnarray}
For $N=3$ and $m =0.01$, we find, for the ground state boson,
\end{minipage}\hspace{.5\columnsep}
\begin{minipage}{0.5\textwidth}
      \begin{center}
         \includegraphics[width=1.1\textwidth]{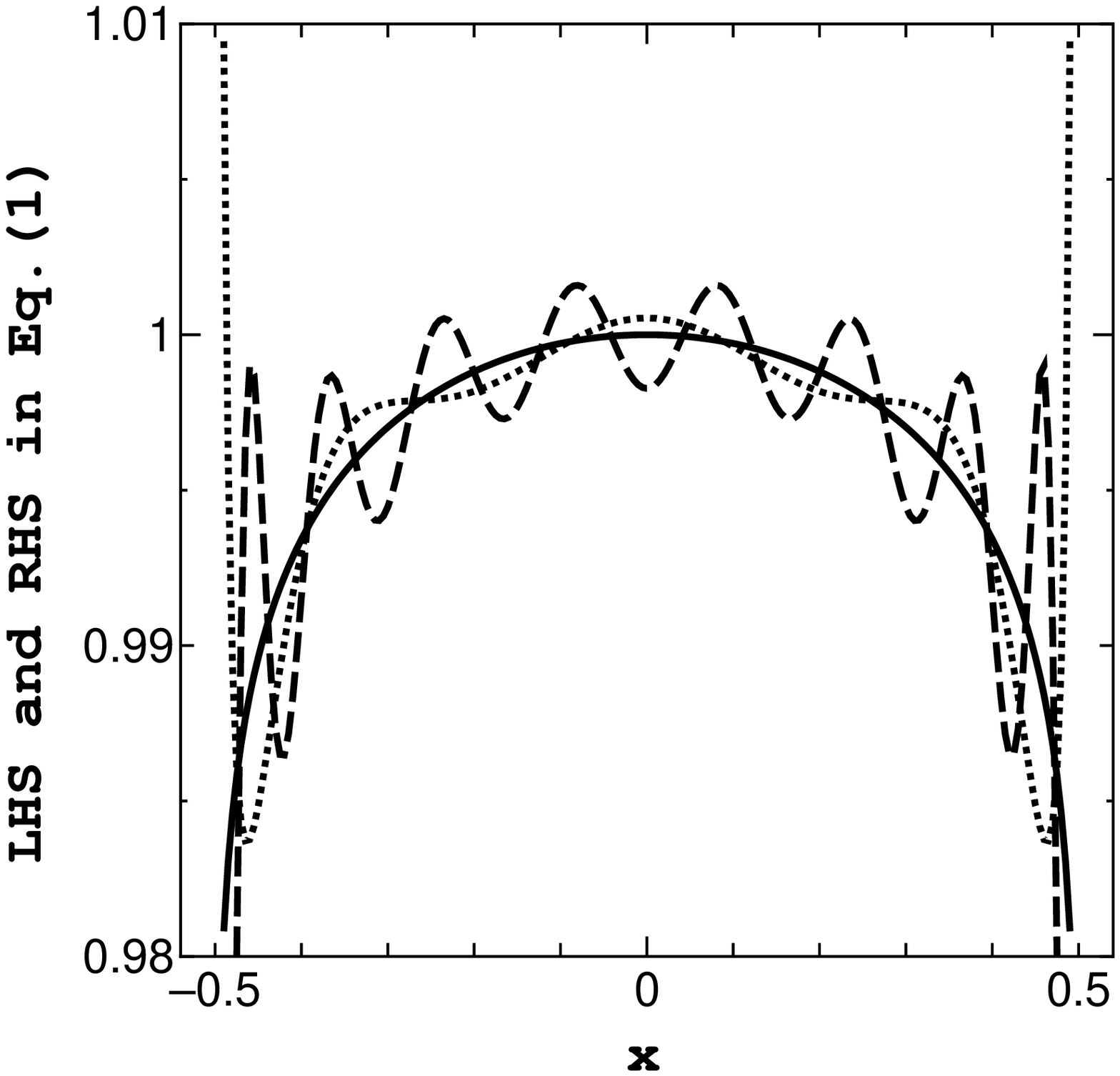}
      \end{center}
\centerline{Fig.~1}
\end{minipage}

\noindent
\begin{eqnarray}
\beta&=&0.00552328,\quad M^2 =0.0366342,\nonumber\\
a_0&=&1,\quad a_1= 0.00203562,\quad
a_2=-0.000579369,\quad 
a_3= 0.000165813.
\label{numeric}
\end{eqnarray}
The values of the LHS and the RHS of Eq. (\ref{thooft}) are shown in 
Fig.~1. In Fig.~1, the solid line represents the LHS and the dotted 
line stands for the RHS. The RHS with $N=9$, which is indicated by 
the dashed line, is exhibited for comparison.
The coincidence of the LHS and the RHS is not inadmissible for small 
values of $x$. On the other hand, for $x\simeq\pm{1/2}$, the 
difference between the LHS and the RHS is not allowable. Only the 
RHS has sharp spikes at the end points. 
This behavior is not changed much even if we improve the order of 
approximation. 

We have to note here that in order to solve the generalized 
eigenvalue problem, 
the norm matrix should be positive definite.
We cannot advance the above procedure beyond $N\simeq 12$, because 
some of the eigenvalues of the norm matrix $\hat N$ become almost 
zero or negative. These facts strongly suggest that the conventional 
basis function is not appropriate basis function for 2 dimensional 
field theories.

\section{New Basis Function}
In order to introduce new basis function, we first notice that there 
are infinite solutions $\beta_1<\beta_2<\cdots$ of Eq. (\ref{mass}) 
in addition to solution $\beta_0$. 
We are easily led to
\begin{eqnarray}
&&0\ll \beta_n-\beta_0-n<{1/2},\nonumber\\
&&0<\beta_n-\beta_k-(n-k)\ll 1,\; n>k.
\end{eqnarray}
The above relations imply that Eq. (\ref{expansion}) never 
incorporates terms like $(1-4x^2)^{\beta_n+j}$ with positive integer 
$n$ and non-negative integer $j$.

We posit that the wave function is given by an infinite series
\begin{eqnarray}
\Phi(x)=\lim_{N\to\infty}\sum_{n=0}^{N}\sum_{j=0}^{N-n} {c_n}^j
   \left(1-4x^2\right)^{\beta_n(m)+j}.
\label{newwave}
\end{eqnarray}
Substituting Eq. (\ref{newwave}) into Eq. (\ref{thooft}), we 
have~\cite{abe99}
\begin{eqnarray} 
0&=&\left.M^2\Phi(x)-\int_{-1/2}^{1/2}dy H(x,y)\Phi(y)
\right|_{1-4x^2=\epsilon}\nonumber\\
 &=&-4\sum_{n=0}^{\infty}{c_n}^0\left
    (m^2-1+\pi\beta_n\cot\pi\beta_n\right)
    \epsilon^{\beta_n-1}\nonumber\\
 &&+\sum_{n=0}^{\infty}\sum_{j=0}^{\infty}
    F_{nj}(m;\beta_n;M;c\cdots )\epsilon^{\beta_n+j}
    \nonumber\\
 &&+\sum_{k=0}^{\infty}
    G_k(m;\beta_0, \beta_1\cdots ;M ;c\cdots )\epsilon^{k}.
\label{series}
\end{eqnarray}
Of course, the first line in 
Eq. (\ref{series}) cancels automatically because of the definition of
$\beta_n$'s. Suppose that we truncate series in 
Eq. (\ref{newwave}) to $O(\epsilon^{\beta_N})$. That is, we set 
${c_n}^j=0$ for $n+j>N$. We demand Eq.  (\ref{series}) to hold up to 
$O(\epsilon^{\beta_N-1})$. Then we have $N(N+3)/2$ non-trivial 
equations.
On the other hand, there are $(N+1)(N+2)/2$ unknown parameters.
The number of parameters is larger than that of non-trivial equations
by 1. Thus, we can solve the equations for ${c_n}^j$ in terms of 
$M^2$.
Another equation of use to us is obtained by multiplying 
both sides of Eq. (\ref{thooft}) by $\Phi (x)$ and integrating them 
over x,
\begin{eqnarray}
M^2\int_{-1/2}^{1/2}dx \left|\Phi(x)\right|^2
&=&\integ\int_{-1/2}^{1/2}dx dy \Phi(x)H(x,y)\Phi(y).
\label{consistency}
\end{eqnarray}

For a given $m$, we put $M^2=M_i^2$.
We can then solve Eq. (\ref{series}) for ${c_n}^j$ in terms of $M_i$.
We thus obtain the $M_i$ dependent truncated wave function, say, 

\noindent
\begin{minipage}{0.5\textwidth}
$\Phi(x;M_i)$. We can calculate a new mass eigenvalue $M_{i+1}$ using 
this wave function as 
\begin{eqnarray}
M_{i+1}^2={<\Phi(M_i)|H|\Phi(M_i)>\over <\Phi(M_i)|\Phi(M_i)>}.
\end{eqnarray}
For $N\le 15$ and $m=0.01$, 
mass $M^2$ converges in 5 iterations. For $0<m<0.5$, we obtain 
$M^2$'s and fit them by polynomials, as follows:
\begin{eqnarray}
M^2(m)=3.6276m + 3.5803{m^2}\nonumber\\
+0.06836{m^3} +O(m^4).
\end{eqnarray}
It should be noted here that the\\
coefficients of $m$ are consistent 
with 
\end{minipage}\hspace{.5\columnsep}
\begin{minipage}{0.5\textwidth}
      \begin{center}
         \includegraphics[width=1.1\textwidth]{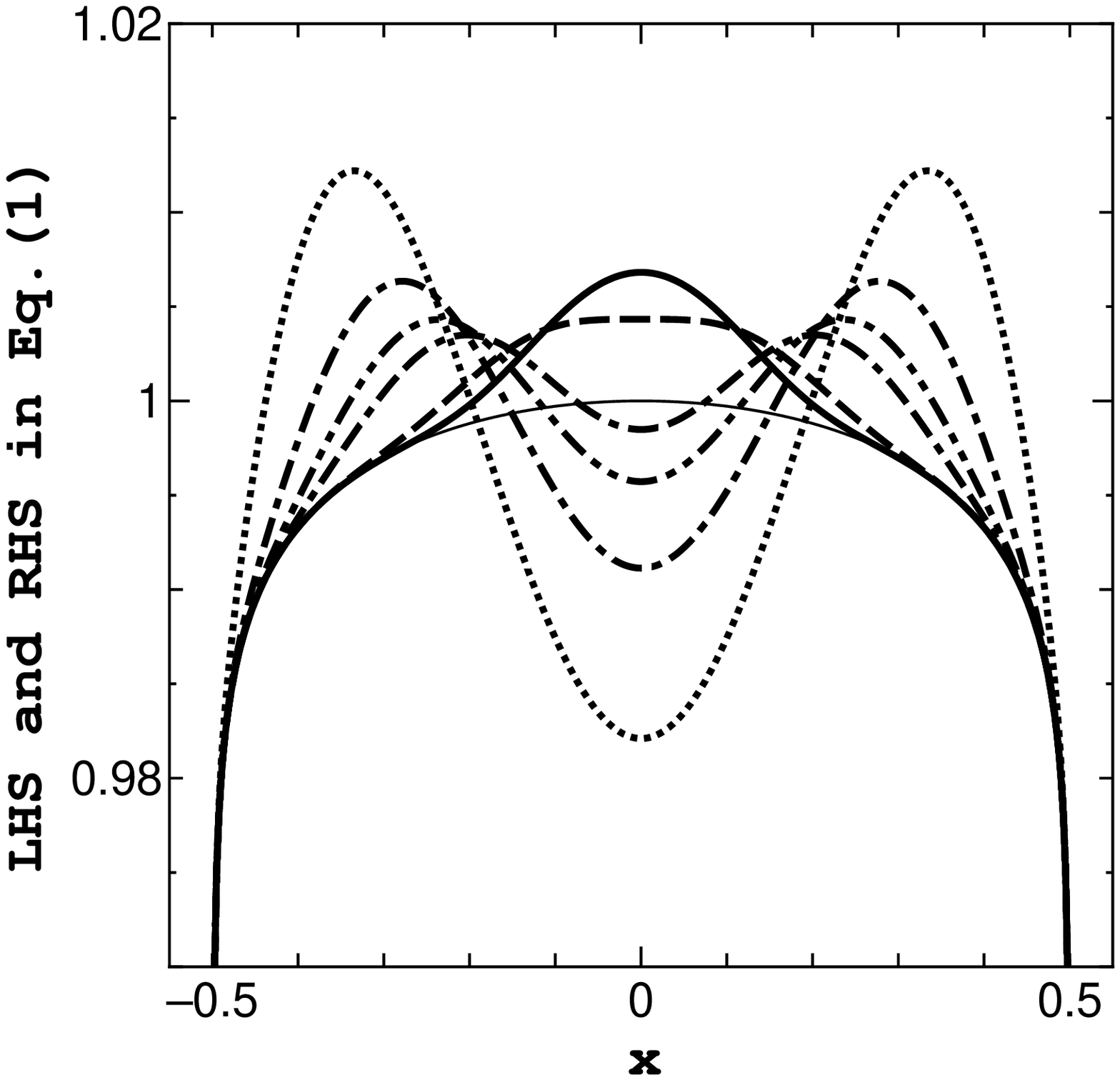}
      \end{center}
\centerline{Fig.~2}
\end{minipage}

\vspace{1.5mm}\noindent
Bergknoff's result.  In order to see the 
efficacy of this new basis function expansion, we show the 
wave functions in Fig.~2. There, the thin solid line represents the 
LHS in Eq. (\ref{thooft}), provided that the wave function was 
approximated by
Eq. (\ref{newwave}) with $N=15$. The dotted line denotes
the RHS with $N=2$, the dot-dashed line exhibits the RHS with $N=3$, 
the dot-dot-dashed line represents the RHS with $N=4$, 
the dot-dash-dashed line stands for the RHS with $N=5$, and the 
dashed line exhibits the RHS with $N=10$. The thick solid 
line indicates the RHS in Eq. (\protect{\ref{thooft}}) with wave 
function given in Eq. (\ref{newwave}) with $N=15$.

\section{Summary and Discussion}
In the previous section we have introduced the new basis function
and calculated the mass eigenvalue of the bound state using the new 
basis function. We have found that 
 (1) the new basis function gives an effective 
approximation of the wave function, and (2) the mass eigenvalues 
are consistent with the results of the precursors. 

It should be noted that Eq.(\ref{newwave}) is, mathematically, 
the most general expansion. This means that there is no room to 
introduce any other additional terms like $d (1-4x^2)^\gamma$ for 
$\gamma\neq\beta_n+j$ with non-negative integers $n$ and $j$. If we 
introduce such terms, then the following equality should hold 
\begin{eqnarray}
0=4d\left(m^2-1+\pi\gamma\cot(\pi\gamma)\right)(1-4x^2)^{\gamma-1}.
\end{eqnarray}
This demands that $d\equiv0$.

\section*{Acknowledgements}
The author would like to thank Professor K. Tanaka and Professor 
G.J. Aubrecht for comments and discussions during the early stage of
this work. He is also grateful to Dr. Harada for useful discussions.
This work was partialy supported by the Grants-in-Aid for Scientific 
Research of Ministry of Education, Science and Culture of Japan 
(No. 10640198).

\section*{References}

\end{document}